\documentstyle[twocolumn,twoside,aps]{revtex}
\input epsf

\renewcommand{\d}{{\rm d}}

\title{Domain formation in transitions with noise and time-dependent bifurcation parameter}
\author{G.D. Lythe\\
{\it
Optique Nonlin\'eaire Th\'eorique,
Universit\'e Libre de Bruxelles CP231,
Bruxelles 1050 Belgium}\\
(received 20 March 1995; revised manuscript received 12 February 1996)\\[2ex]
\hspace{-0.8em}\parbox{16.7cm}{
 The characteristic size for spatial structure, that
emerges when the bifurcation parameter 
in model partial differential equations is
slowly increased through its critical value,
depends logarithmically on the
size of added noise. Numerics and analysis
 are presented for the real Ginzburg--Landau
 and Swift--Hohenberg equations.
}
}
\begin{document}
\vskip 2cm
\maketitle

Many physical systems undergo a transition from a spatially uniform
state to one of lower symmetry. 
Classical examples are the formation of magnetic domains
 and the Rayleigh-Benard instability [1].
Such systems are commonly modeled
by a simple differential equation, having a bifurcation parameter
with a critical value at which the spatially uniform
state loses stability.
Noise is often assumed to
provide the initial symmetry-breaking perturbation permitting
 the system to choose one of
the available lower-symmetry states,
 but is not often explicitly included in mathematical models.
However, when the bifurcation parameter
is slowly increased through its critical value
 it is necessary to consider noise explicitly.

The phenomenon of delayed bifurcation and its sensitivity to noise
has been reported in the case of non-autonomous stochastic
ordinary differential equations [2]; here the corresponding
phenomenon is examined in partial differential equations.
A characteristic length for the spatial pattern is demonstrated from 
a stochastic partial differential  equation (SPDE),
supported by numerical simulations.
Noise is added in such a way that it has no correlation length of its own
 (white in space and time) and a finite difference algorithm is used whose
continuum limit is an SPDE.

The mathematical description of transitions
is in terms of a space-dependent order 
parameter $Y$ and a bifurcation parameter $g$.
Because it is the simplest model with the essential features,
the real Ginzburg--Landau equation (GL)  is considered first.
Results are also presented for the Swift--Hohenberg equation (SH),
that is more explicitly designed to model Rayleigh-Benard convection.

When the bifurcation parameter $g$ is constant the following
is found. For $g<0$, in both GL and SH, the solution with
$Y$ everywhere $0$ is stable.
In GL for $g>0$ one sees a pattern of regions where $Y$
is positive and regions where $Y$ is negative (domains)
separated by narrow transition layers. 
 In SH for $g>0$ there is a structure resembling a pattern of parallel rolls, 
interrupted by defects.

When $g$ is {\it slowly increased through} $0$ 
{\it in the presence of noise}
a characteristic length is produced as follows. The field
$Y$ remains everywhere small until well after $g$ passes through $0$.
At $g\simeq g_c$, where
\begin{eqnarray}
g_c=\sqrt{2\mu|\ln\epsilon |},
\end{eqnarray}
 $\mu$ is the rate of increase of $g$ and
$\epsilon$ is the amplitude of the noise,
$Y$ at last becomes ${\cal O}{(1)}$ 
and the spatial pattern present is frozen in by the nonlinearity.
Thereafter one observes spatial structure with 
characteristic size proportional to
$({|\ln\epsilon|/\mu})^{{1\over 4}}$.
In GL this length is the typical size of the domains;
in SH it is the typical distance betwen defects.

The results reported here were obtained by solving
SPDEs [3] of the following dimensionless
form for stochastic processes $Y$ depending
on $x$ and $t$:
\begin{eqnarray}\d Y = [g(t)Y-Y^3 + {\cal L} Y]\d t + \epsilon \d W.\end{eqnarray}
The equations were solved as initial value problems, with 
$g(t)=\mu t$ slowly  increased from $-1$ to $1$.
Here $Y:[0,L]^m\times[-{1\over\mu},{1\over\mu}]\times\Omega \to \cal{R}$,
$\Omega$ is a probability space and $W$ is the Brownian sheet [4],
the generalisation of the Wiener process (standard Brownian motion)
to processes dependent on both space and time.
Periodic boundaries in $x$ are used so that any 
spatial structure is not a boundary effect.
The constants $\mu$, $\epsilon$ and ${1\over L}$
 are all $\ll 1$. Results are reported for
${\cal L}=\Delta$ (GL)
and
${\cal L} = -(1+\Delta)^2$ (SH) where
$\Delta=\sum_{i=1}^m{\partial^2\over\partial x_i^2}$, the Laplacian
in ${\cal R}^m$.

In the first order finite difference algorithm for numerical 
realisations of the lattice version of (2),
$y_{t+\Delta t}(i)$ is generated from $y_t(i)$ as follows:
\begin{eqnarray}
y_{t+\Delta t}(i) = &y_t(i)+[\mu t y_t(i)-y_t^3(i)
+\tilde{\cal{L}}
y_t(i) ]\Delta t\nonumber\\
&+ \epsilon\, \left(\Delta x\right)^{-{m\over 2}}\,
 n_t(i)\, \sqrt{\Delta t}.
\end{eqnarray}
In (3), $y_t(i)$ is numerical approximation to
the value of $Y$ at site $i$ at time $t$ and
$\tilde {\cal{L}}$ is the discrete version of ${\cal{L}}$. 
The $n_t(i)$  are
 Gaussian random variables with unit variance, independent
of each other, of the values at other sites, and of the values
at other times.
  
It is also possible to introduce multiplicative noise,
for example to make $g$ a random function of space and time [5,6].
The effect in that case is proportional to the magnitude of
the noise and is thus less dramatic at small intensities than that
of additive noise.

\pagebreak[2]
The timing of the emergence of spatial structure
can be understood by deriving the stochastic ordinary
 differential equation for the
most unstable Fourier mode, which is of the form [7]
\pagebreak[3]
\begin{eqnarray}
\d y = [g(t)y-y^3]\d t + \epsilon \d w,
\end{eqnarray}
where $w$ is the Wiener process.
Trajectories of (4) remain close to $y=0$ until well after $g=0$, 
then jump abruptly towards one of the new attractors (Figure 1).
The value of $g$ at the
 jump can be determined by solving the 
linearised version ; for $\mu\ll 1$ it
 is a random variable with mean approximately $g_c$
 and standard deviation proportional to $\mu$ [8].

\begin{figure}
\epsfbox{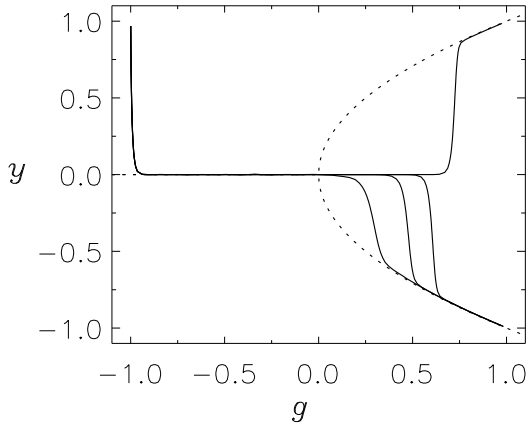}
\caption{
{\it Dynamic pitchfork bifurcation with noise.} 
The dotted lines are the loci of stable fixed points
of  $\dot y=gy-y^3$ as a function of $g$.
The solid lines are solutions of the non-autonomous 
SDE (4) with $g=\mu t$,  for noise levels
$\epsilon=10^{-3}$, $10^{-6}$, $10^{-9}$, $10^{-12}$.
(In each case $\mu=0.01$ and  the
 initial condition is $y=1.0$ at $g=-1.0$.)
}
\label{pffig}
\end{figure}

The Ginzburg-Landau equation is
a simple model of a spatially extended system where
a uniform state loses stability 
to a collection of non-symmetric states.
When $g$ is fixed and positive in this equation,
 a pattern of domains  is found.
 In each domain, $Y$ is close either to $\sqrt{g}$ or to $-\sqrt{g}$. 
The gradual merging of domains on extremely long timescales [9]
is not the subject of this paper; here the focus is on
how the domains are formed by a slow increase of the bifurcation
 parameter through $0$.
An example  is depicted in Figure 2:
a pattern of domains emerges 
when $Y$ is everywhere small and is frozen in at $g\simeq g_c$.
When $Y$ is small
an excellent approximation to the correlation function,
$c(x)=\big<Y_t(v)Y_t(x+v)\big>$, can be calculated
from the solution of the linearised version 
of (2) (that is, without the cubic term).
The correlation length at $g=g_c$ becomes the characteristic length for
spatial structure after $g=g_c$.

For GL,
the solution of the linearised version of (2) is:
\begin{eqnarray}
Y_t(x) = \int_{[0,L]^m}G({\scriptstyle t,-{1\over\mu},x,v})f(v)\d v +
\nonumber\\
\epsilon\int_{-{1\over\mu}}^t\int_{[0,L]^m}G({\scriptstyle t,s,x,v})\d v\d W_s(v),
\end{eqnarray}
\begin{figure}
\epsfbox{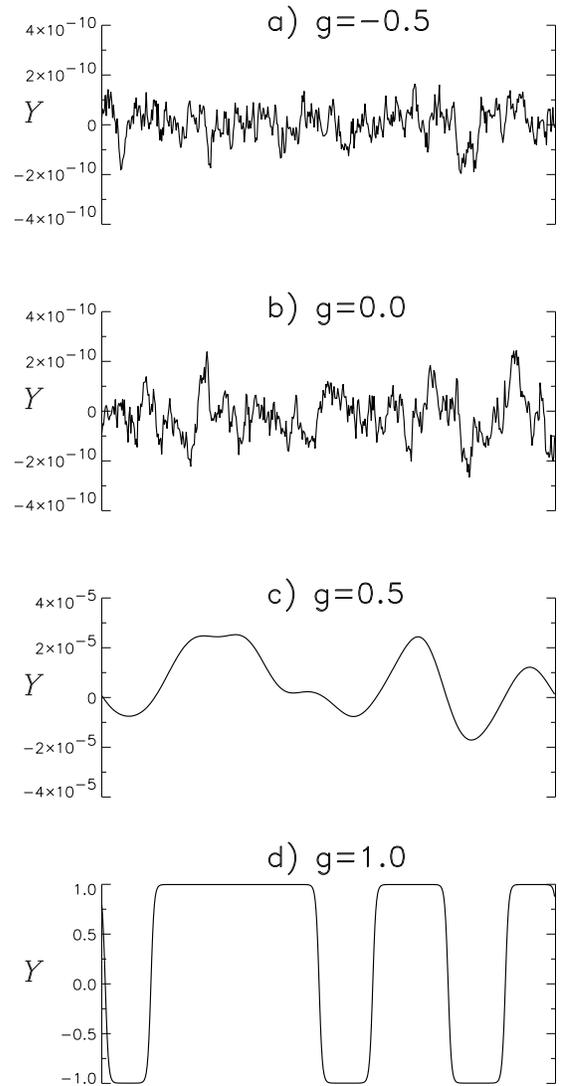}
\caption{
{\it Dynamic transition, GL,
 one space dimension.} 
Four configurations, $Y_t(x)$, are shown from
one numerically-generated realisation 
of the SPDE
(note the different vertical scales).  Nonlinear
terms become important when $g\simeq 0.64$; their effect is
to freeze in the spatial structure.
 ($L=300$, $\mu=0.01$, $\epsilon=10^{-10}$.)}
\label{oned}
\end{figure}
\pagebreak[3]
where $G({\scriptstyle t,s,x,v})= $
%\begin{equation}
$$
{\scriptstyle [4\pi(t-s)]}^{-{m\over2}}
\exp\left({\scriptstyle-{\,\,(x-v)^2\over4(t-s)}-\mu(t^2-s^2)}\right)
$$
%\end{equation}
with $x-v$ understood modulo $[0,L]^m$.
The first term, dependent on the initial data $f(x)$, relaxes quickly to 
very small values and remains negligible if $2\mu|\ln\epsilon|<1$.
\pagebreak[3]
The correlation function is therefore obtained
from the second integral in (5).
The mean  of the product of two such stochastic integrals
 is an ordinary integral [4]. 
 Performing the integration over space [7], assuming that
$L>({8\over\mu})^{{1\over 2}}$, gives
\begin{eqnarray}c(x)=
\epsilon^2\int_0^t
{{\rm e}^{\mu(t^2-s^2)}{\rm e}^{{-x^2\over8(t-s)}}\over
[8\pi(t-s)]^{{m\over2}}}\d s.\end{eqnarray}
Before $g$ approaches $0$, the correlation function
 differs by only
${\cal{O}}({\mu\over g^2})$ from its static ($g=$constant) form [7]; it
 remains well-behaved as $g$ passes through $0$ and,
 for $g > \sqrt{\mu}$, is well approximated by:
\begin{eqnarray}c(x) \simeq
{\epsilon^2{\rm e}^{\mu t^2} \over ({8\mu t})^{{m\over 2}}}
{\rm e}^{-{x^2\over 8t}}.\end{eqnarray}

For ${1\over\sqrt{\mu}}<g<g_c$, typical values of $Y(x)$ increase exponentially fast and the correlation length is proportional to $\sqrt{t}$. Effectively
 noise acts for $g\le\sqrt{\mu}$ to provide an initial condition for the 
subsequent evolution.
At a value of $g$ that is a random variable with mean
$g\simeq g_c$ and standard deviation proportional to $\mu$,
 the cubic nonlinearity  becomes important.
Its effect is to freeze in the spatial structure;
no perceptible changes occur between $g=g_c$ and $g=1$.

In one space dimension
 it is possible to put the scenario just described
to quantitative test
by producing numerous realisations like that of Fig.2 and recording
 the number of times $Y$ crosses upwards through
$0$ in the domain $[0,L]$ at $g=1$.
 In Fig.3 the average number of upcrossings
is displayed as a function of the sweep rate $\mu$.
  The solid line is the expected number of upcrossings of zero,
\begin{eqnarray}r={L\over2\pi}\left({-c''(0)\over c(0)}\right)^{\frac12}
={L\over4\pi}
\left({\mu\over 2|\ln\epsilon|}\right)^{{1\over 4}},\end{eqnarray}
 for a field 
{\it with correlation function (7) at $g=g_c$} [10]. The hypothesis
that the spatial pattern does not change after $g=g_c$ is
succesful.
\begin{figure}
\epsfbox{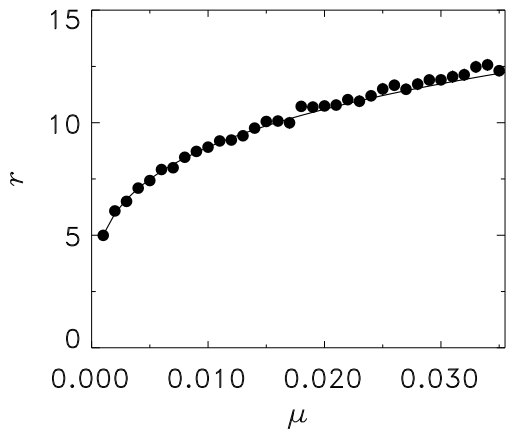}
\caption{
{\it Number of zero crossings after a dynamic transition}.
The dots are the mean number of upcrossings of $0$ at $g=1$ in
numerical realisations of GL in one space dimension.
The solid line is the prediction based on the assumption that the 
correlation function (7) is valid until $g=g_c$, after
which time the spatial pattern does not change.
($\epsilon=10^{-4}$  and $L=800$.)}
\label{test}
\end{figure}

In one space dimension, the solution of the SPDE (2) is a stochastic
process with values in a space of continuous 
functions [3,12,13]. That is, for
fixed $\omega\in\Omega$ and $t\in[-{1\over\mu},{1\over\mu}]$,
 one obtains a configuration, $Y_t(x)$, 
 that is a continuous function of $x$. This can
be pictured as the shape of a string at time $t$ that is 
constantly subject to small random impulses all along its length.
 In more than one space dimension, however, the $Y_t(x)$
 are not necessarily continuous functions but only distributions [3,12].
 Typically the correlation function
$c(x)$ diverges at $x=0$.
In the dynamic equations studied here,
however, the divergent part does not grow exponentially for $g>0$, and by
 $g= g_c$ it
is only apparent on extremely small scales, beyond the resolution of
any feasible finite difference algorithm.
Figure 4 depicts configurations  at $g=1$
 from realisations of (2) in two space dimensions.
In Figures 4(a) and 4(b) (GL) one sees that a faster rate of increase of $g$
results in a smaller average domain size.
The SPDEs were simulated on a grid of $512\times 512$ points 
with second order timestepping [13].

%\onecolumn
\begin{figure}
\centerline{
\epsfysize=4cm
\epsffile{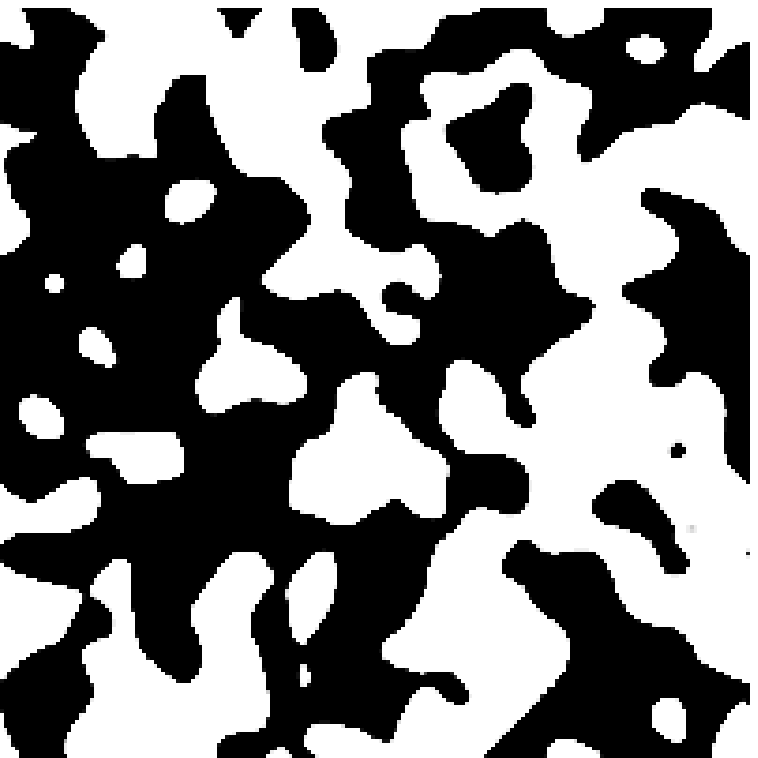}
\ \ \ 
\epsfysize=4cm
\epsffile{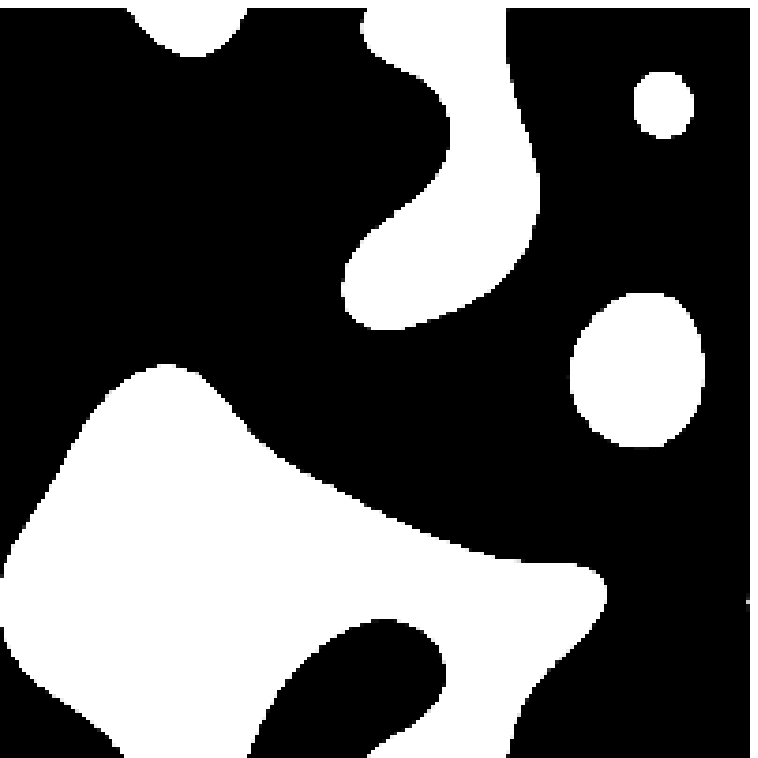}}
\vspace{0.5cm}
\centerline{
\epsfysize=4cm
\epsffile{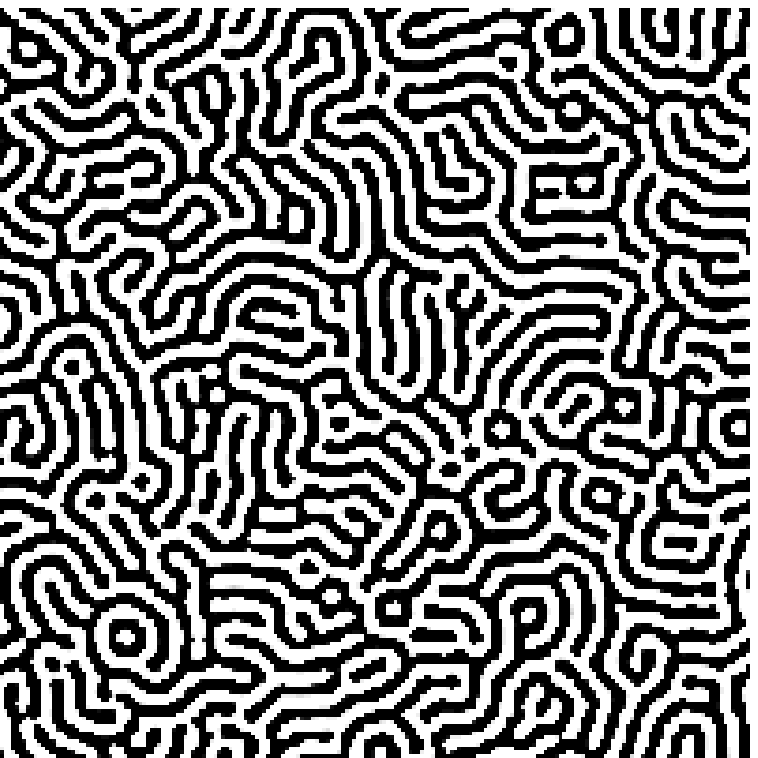}
\ \ \ 
\epsfysize=4cm
\epsffile{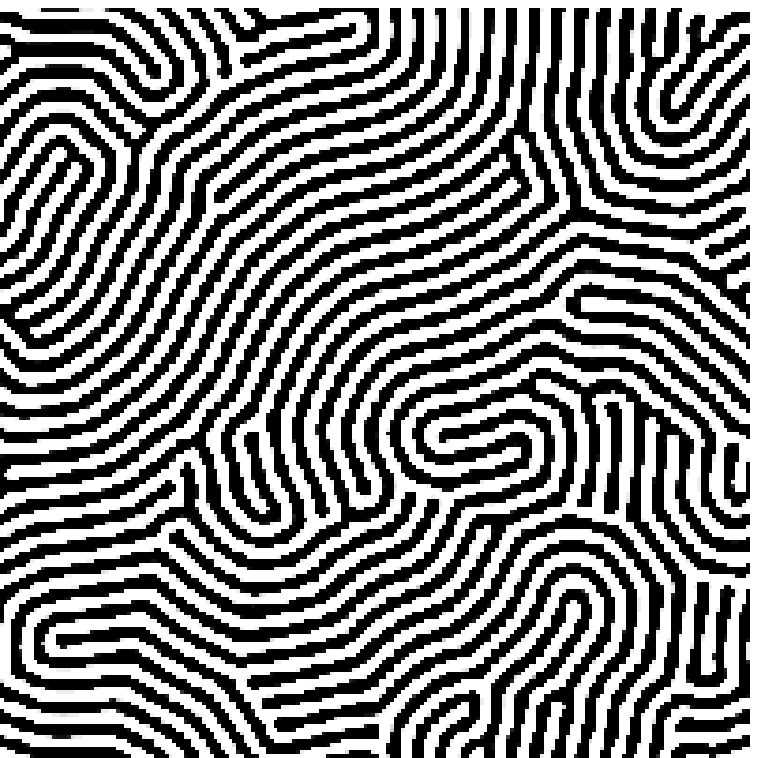}}
\vspace{1cm}
\caption{
{\it Two-dimensional pattern at $g=1$:
smaller $\mu$ means larger characteristic length.}
In black regions $Y<0$;
in white or grey regions $Y>0$. 
 In GL, (a) and (b),
the typical domain size decreases with $\mu$
, the rate of increase of $g$. In SH, (c) and (d),
where there is a short-range structure resembling parallel rolls,
the effect of reducing $\mu$
 is to reduce the number of defects.
 4(a): GL,  $L=300$, $\epsilon=10^{-5}$,
 $\mu=0.03$.
 4(b): GL, $L=300$, $\epsilon=10^{-5}$,
 $\mu=0.003$.
 4(c): SH, $L=200$, $\epsilon=10^{-5}$,
 $\mu=0.01$.
 4(d): SH, $L=200$, $\epsilon=10^{-5}$,
 $\mu=0.001$.
}
\end{figure}
%\twocolumn

The essential difference between the
Swift-Hohenberg  and Ginzburg--Landau models
 is that the first spatial Fourier
mode to become unstable has $k=1$ rather than 
\onecolumn
\noindent
\parbox{8.5cm}{
$k=0$. 
Hence there is a preferred small-scale
pattern that resembles the parallel rolls seen in experiments.
However,  there is no preferred orientation of the roll pattern
and when the correlation length is smaller than the system size,
 many defects are found, separating regions where
 the rolls have different orientations. When $g$ is increased through $0$,
the number of defects resulting decreases when $\mu$ decreases --
 Fig.4(c) and (d). Here
 a grid of $300\times 300$ points was used
with first order timestepping.

A notable feature of dynamic bifurcations and dynamic transitions is that the evolution for $g>0$ is independent of the initial conditions (provided they are such that that the system descends into the noise). Noise acts, near $g=0$,
}
\ \ \ \ 
\parbox{8.5cm}{
to wipe out the memory of the system and to provide an initial condition for the subsequent evolution. The correlation 
function (7) is, for example, a natural initial condition for
studying the dynamics
of defects and phase separation  because it emerges
 from a slow increase to supercritical
 of the bifurcation parameter in the presence of 
space-time noise, mimicking an idealised experimental situation.

In summary, dynamic transitions are analysed 
 in models of spatially extended systems with white noise.
The correlation length that emerges from the noise during a slow sweep
past $g=0$ is frozen in by the nonlinearity as a
characteristic length
 proportional to $({|\ln\epsilon|/\mu})^{{1\over 4}}$ 
where $\mu$ is the rate of increase of the bifurcation parameter and
$\epsilon$ is the amplitude of the noise.
}

\vskip 1cm
{\hskip 4cm }\hrulefill {\hskip 4cm }
\vskip 1cm
\parbox{8.5cm}{
%\noindent{\bf References}

\noindent [1] D.J. Scalapino, M. Sears and R.A. Ferrell,
        Phys. Rev. B {\bf 6} 3409 (1972);
        Robert Graham, Phys. Rev. A {\bf 10} 1762 (1974);
        M.C. Cross and P.C. Hohenberg,
            {Rev. Mod. Phys.} {\bf 92} 851 (1993).

\noindent [2]
{C. W. Meyer, G. {Ahlers} and D. S. {Cannell}},
             {Phys. Rev. A} {\bf 44} 2514 ({1991}),
Jorge Vi\~ nals, Hao-Wen Xi and J.D. Gunton,
             {\it ibid.} {\bf 46} 918 ({1992}),
P.C. Hohenberg and J.B. Swift
             {\it ibid.} {\bf 46} 4773 ({1992)},
Walter Zimmermann, Markus Seesselberg and Francesco Petruccione
             {\it ibid.} {\bf 48} 2699 ({1993}).

\noindent [3]
{{J.B. Walsh}, in:
            {\sl Ecole d'\'et\'e de probabilit\'es de St-Flour XIV}
            edited by P.L. Hennequin }(Springer, Berlin, 1986) pp266--439;
{{G.} {Da Prato} and {J.} {Zabczyk}},
    {\sl Stochastic Equations in Infinite Dimensions}
    {(Cambridge University Press, Cambridge, 1992)}.

\noindent [4]
If $f_1(x,t)$ and $f_2(x,t)$ are continuous functions on
${\cal{D}}\times\cal{T}$, where ${\cal{D}}\in{\cal R}^m$
and ${\cal{T}}\in\cal{R}$, then 
$I_1=\int_{\cal{T}}\int_{\cal{D}}
f_1\d x\d W$ and $I_2=\int_{\cal{T}}\int_{{\cal{D}}}f_2\d x\d W$ are Gaussian
random variables with $<I_1>=0$, 
$<I_2>=0$
and $<I_1 I_2>=
\int_{\cal{T}}\int_{\cal{D}}f_1(x,t)f_2(x,t)\d x\d t.$

\noindent [5]
{{C. R.} {Doering}},
           {Phys. Lett. A} {\bf 122} 133 {(1987)};
{{A.} {Becker} and {Lorenz} {Kramer}},
            {Phys. Rev. Lett.} {\bf 73} 955 ({1994}).

\noindent [6]
{J.} {Garc\'\i a-Ojalvo}  and {J.M.} {Sancho},
            {Phys. Rev. E} {\bf 49} 2769 (1994);
{L.} {Ram\'\i rez-Piscina}, {A.}{Hern\'andez-Machado} and {J.M.} {Sancho},
            {Phys. Rev.B} {\bf 48} 119 (1994);
}
\ \ \ \ 
\parbox{8.5cm}{
{J.} {Garc\'\i a-Ojalvo}, {A.}{Hern\'andez-Machado} and {J.M.} {Sancho},
            {Phys. Rev. Lett.} {\bf 71} 1542 (1994).

\noindent [7]
{{G.D.} {Lythe}},
            {in: Stochastic Partial Differential Equations},
                edited by {Alison} {Etheridge} ({Cambridge University Press},
                {Cambridge}, {1994}).

\noindent [8]
{{M.C.} {Torrent} and {M.} {San Miguel}},
           {Phys. Rev. A} {\bf 38} 245 ({1988});
{N.G.} {Stocks}, {R.} {Mannella} and {P.V.E.}{McClintock},
            {{\it ibid.}} {\bf 40} 5361 (1989);
{J.W. Swift}, {P.C. Hohenberg}
 and {Guenter Ahlers},
            {\it ibid.} {\bf 43} 6572 (1991);
{G.D. Lythe and M.R.E. Proctor}, 
 {Phys. Rev. E. } {\bf 47} {3122-3127} {(1993)};
{Kalvis Jansons and Grant Lythe}, submitted to J. Stat. Phys. (1996).

\noindent [9]
{{J.} {Carr} and {R.} {Pego}},
            {Proc. R. Soc. London A}{436} 569 ({1992}); 
{{J.} {Carr} and {R.L.} {Pego}},
            {Comm. Pure Appl. Math.}{42} 523 ({1989}).

\noindent [10]
K. Ito 
J. Math. Kyoto Univ. {\bf 3-2} 207 (1964);
{{R. J.} {Adler}},
            {\sl The Geometry of Random Fields}
            ({Wiley}, {Chichester}, 1981).

\noindent [11]
{T. Funaki},
            Nagoya Math. J. {\bf 89} 129 (1983);
{I. Gy\"ongy} and {E. Pardoux}, 
            Probab. Theory Relat. Fields {\bf 94} 413 (1993).

\noindent [12]
{C. R. Doering},
           Comm. Math. Phys. {\bf 109} 537 (1987).

\noindent [13]
Grant Lythe, {\sl Stochastic slow-fast dynamics} (PhD 
Thesis, University of Cambridge, 1994);
{Peter E.} {Kloeden} and {Eckhard} {Platen},
 {\sl Numerical Solution of Stochastic Differential Equations}
 (Springer, Berlin, 1992)
}

\end{document}